\documentclass{article}
\usepackage{amssymb,amsmath,amsthm,stmaryrd,mathrsfs}

\newfont{\Fr}{eufm10}

\newcommand{\bb}{\mathbb}

\usepackage{hyperref}
\begin{document}

\title{Geometric construction of the Quantum Hall Effect in all even dimensions}
\author{Guowu Meng\\ \small{\it Department of Mathematics}\\ \small{\it Hong Kong University of Science and Technology}\\
\small{\it Clear Water Bay, Kowloon, Hong Kong}\\
\small{Email: mameng@ust.hk}}

\maketitle

\begin{abstract}
The Quantum Hall Effects in all even dimensions are uniformly
constructed. Contrary to some recent accounts in the literature,
the existence of Quantum Hall Effects does not {\it crucially}
depend on the existence of division algebras. For QHE on flat
space of even dimensions, both the Hamiltonians and the ground
state wave-functions for a single particle are explicitly
described. This explicit description immediately tells us that QHE
on a higher even dimensional flat space shares the common features
such as incompressibility with QHE on plane.
\end{abstract}


\section{Introduction}
Recently there is a flurry on a generalization of the Quantum Hall
Effect (QHE) to four dimensional flat space \cite{BHTZ03, CL, Sp,
Chen, EP, BCHTZ, KN, ZH02,MF02, HOU02}, inspired by a paper
\cite{ZH01} by Zhang and Hu. In view of its significance to the
condensed matter physics and to the fundamental physics (see the
conclusion and the introduction in \cite{ZH01}), it is worth the
effort to have a closer examination of this generalization of the
QHE.

In their search for QHE on four-space (i.e., ${\bb R}^4$), Zhang
and Hu follow Haldane's approach of the QHE problem in
\cite{haldane}. There are two steps in this approach: 1) study the
quantum mechanics problem of a single charged particle under the
influence of a natural background magnetic field of strength $I$
on the sphere of radius $R$; 2) map the sphere to the flat
Euclidean space by the standard stereographical mapping and then
take the thermodynamic limit as both $I$ and $R$ go to infinity
while keeping $I/R^2$ constant to recover QHE on the plane. Put it
differently, QHE on two-sphere devised by Haldane is just a family
of spherical approximations to QHE on two-space (i.e., ${\bb
R}^2$), and the strategy adopted in \cite{ZH01} is to generalize
the spherical models of Haldane to dimension four (using an
earlier work of C. N. Yang \cite{yang}) and then take the
thermodynamic limit to obtain QHE on four-space.

As a matter of fact, the natural generalization of QHE on
two-sphere goes beyond dimension four. This observation was also
independently made (but not carried out) by M. Fabinger in
\cite{MF02} from the point of view of fuzzy spheres \cite{SR01}.

It is well known that, to understand the QHE, a key step is to
understand the Hamiltonian and the ground state wave-functions for
a single particle on flat space. However, to our best
understanding of the references cited above, this very important
question for QHE on four-space has not been explicitly addressed
so far. Our answer to this question for QHE on higher dimensional
flat spaces turns out to be rather simple mathematically. This
simplicity immediately tells us that QHE on higher dimensional
flat space shares the common features such as incompressibility
with QHE on two-space.

\vskip 10pt I would like to thank the referees and the Board
Member for their helpful suggestions. I would like to thank J. S.
Li, X. R. Wang and S.C. Zhang for helpful discussions. Especially
I would like to thank Y.S. Wu for very helpful and very
constructive discussions on the physical meaning of the main
result of this paper. This work is supported by the Hong Kong
Research Grants Council under the RGC project HKUST6161/97P.

\section{QHE on even dimensional spheres}
Here we formulate the quantum mechanics model\footnote{Before
going into the mathematical details, I would like to point out
that, physically, it is the quantum mechanics of a charged
particle in $2n$-dimensional sphere under the influence of a
canonical background $\hbox{Spin}(2n)$-gauge field. For more
explanation, please consult appendix A. } for QHE problem on
even-spheres in clean geometric language. The approach in
\cite{ZH01}, where the ground state wave-functions are given first
and the Hamiltonian is derived later, while works in dimension
four, does not seem to work in higher even dimensions. Our
approach starts with the Hamiltonian and is more straightforward
and works in any even dimension. However, the discussion here may
not be really new in a broad sense, and it could be known long
time ago to mathematicians in the context of representation theory
of compact Lie groups. In any case, similar discussions in more
general settings appeared (partially or fully) in the mathematical
physics literature repeatedly in the past \cite{MS89, LAND92,
MT96}.

Following \cite{ZH01}, a point $X_i$ on $\hbox{S}^{2n}(R)$ (the
$2n$-sphere centered at the origin with radius $R$) can be
described by dimensionless vector coordinates $x_i=X_i/R$, with
$i=1,2,\ldots, 2n+1$, which satisfy $x_i^2=1$. Now,
$\hbox{S}^{2n}\equiv \hbox{S}^{2n}(1)$ is the homogeneous space
$\hbox{Spin}(2n+1)/\hbox{Spin}(2n)$, and the principal bundle
\begin{eqnarray}\label{pb} \hbox{Spin}(2n)\rightarrow \hbox{Spin}(2n+1)\rightarrow \hbox{S}^{2n}
\end{eqnarray}
has a canonical connection
\begin{eqnarray} A=Pr_{so(2n)}(g^{-1}dg),
\end{eqnarray}
where $g^{-1}dg$ is the Cartan-Maurer form on $\hbox{Spin}(2n+1)$
and $Pr_{so(2n)}$ is the orthogonal projection onto the Lie
algebra of $\hbox{Spin}(2n)$. Let $\Delta$ be the fundamental spin
representation of $\hbox{Spin}(2n+1)$, then $\Delta=\Delta^+\oplus
\Delta^-$ as representations of $\hbox{Spin}(2n)$, where
$\Delta^\pm$ are the positive/negative spin representation of
$\hbox{Spin}(2n)$. The highest weight state of $\Delta^\pm$ is
$\left |\underbrace{\hbox{${1\over 2}\cdots {1\over
2}$}}_{n-1}{\pm1\over 2}\right >$. Let $I$ be a positive half
integer, $\Delta^+_{I}$ be irreducible representation of
$\hbox{Spin}(2n)$ with highest weight state $\left
|\underbrace{I\cdots I}_n\right >$. (In general $\Delta_I^+$ is an
irreducible component of the $2I$-fold symmetric tensor product of
$\Delta^+$) Form the complex vector bundle $\xi_I$:
$\hbox{Spin}(2n+1)\times_{\hbox{Spin}(2n)}\Delta^+_I\rightarrow
\hbox{S}^{2n}$, then $\xi_I$ has an induced canonical
$\hbox{Spin}(2n)$-connection $A_I$. The quantum mechanics problem
is the study of a charged particle under the presence of
background magnetic potential $A_I$, so the
Hamiltonian\footnote{Here is a remark for QHE on four-sphere. Note
that $\hbox{Spin}(4)=\hbox{SU}(2)\times \hbox{SU}(2)$, so the
background gauge field splits into two components. However, since
the particle is neutral with respect to the 2nd component gauge
field, physically, the particle only sees $\hbox{SU}(2)$ --- a
component of $\hbox{Spin}(4)$. So the effective principal bundle
used here is just a $\hbox{SU}(2)$-bundle which can be seen to be
precisely the Hopf bundle used in \cite{ZH01}. (These observations
have already appeared in a series of papers of Y.S. Wu and his
collaborators which are published in some Chinese journals in the
seventies.) The advantage of our construction is that the
$\hbox{Spin}(5)$(not $\hbox{SO}(5)$) symmetry of the system is
manifest from the very beginning. Please compare with
\cite{ZH01}.} is
\begin{eqnarray}\label{E:hamil}
\hat H={\hbar^2\over 2MR^2}d^\dag_{A_I} d_{A_I},
\end{eqnarray}
where $d_{A_I}$ is the covariant derivative:
$\Gamma(\xi_I)\rightarrow \Gamma(\xi_I\otimes T^*\hbox{S}^{2n}),$
and $d^\dag_{A_I}$ is the formal adjoint of $d_{A_I}$.

To compute the spectrum of $\hat H$ in equation (\ref{E:hamil}),
we note that
\begin{eqnarray}\label{E:C}
{d^\dag_{A_I}}
{d_{A_I}}=c_2\left(\hbox{Spin}(2n+1)\right)-c_2(\hbox{Spin}(2n),
\Delta_I^+)
\end{eqnarray}
where $c_2(\hbox{Spin}(2n+1))$ is the quadratic Casimir operator
of $\hbox{Spin}(2n+1)$ and $c_2(\hbox{Spin}(2n), \Delta_I^+)$ is
the value of quadratic Casimir operator of $\hbox{Spin}(2n)$ on
$\Delta_I^+$. Therefore we have
\begin{eqnarray}\label{H}
\hat H={\hbar^2\over
2MR^2}[c_2(\hbox{Spin}(2n+1))-c_2(\hbox{Spin}(2n), \Delta_I^+)]
\end{eqnarray}
Note that equations (\ref{E:hamil}) and (\ref{H}) have appeared in
\cite{MS89,LAND92,MT96} in slightly different
form. (see Appendix A for more details)

The Hilbert space of this quantum system is the space of square
integrable sections of $\xi_I$ and it decomposes into the direct
sum of the eigenspaces of $\hat H$. These energy eigenspaces,
indexed by integer $q\ge 0$, are all irreducible representation
spaces of $\hbox{Spin}(2n+1)$. The $q$-th energy eigenspace ${\cal
H}_I(q)$ is labelled by its highest weight state $\left
|(q+I)\underbrace{I\cdots I}_{n-1}\right >$, and the corresponding
eigenvalue is $E(q)=\frac{\hbar^2}{2MR^2}\left[2I(q+{n\over
2})+q(q+2n-1)\right]$. The ground state,  which is the lowest
$\hbox{Spin}(2n+1)$ level for a given $I$, is obtained by setting
$q=0$, and is $d_0\equiv\prod_{1\le i\le j\le n}\left(1+{2I\over
2n+1-i-j}\right)$-fold degenerate. (All of these are standard
results in mathematics and can be found in a textbook on group
representations, for example \cite{DZ73}) Therefore, $I$ plays the
role of the magnetic flux, while $q$ plays the role of the Landau
level index. States with $q>0$ are separated from the ground state
by a finite energy gap. Note that in the limit
$I\rightarrow\infty$ and $R\rightarrow\infty$ while keeping
$l_0\equiv{R\over \sqrt {2I}}$ and $q$ constant,
\begin{eqnarray}\label{E:es}
E(q)\rightarrow \frac{\hbar^2}{2Ml_0^2}(q+{n\over 2}),
\end{eqnarray}
and the single particle energy spacing is finite. This can also be
seen from the thermodynamic limit of $\hat H$ in the below.

\section{QHE on even dimensional flat spaces}
Here we shall see that $\hat H$ in equation (\ref{E:hamil}) is
rather simple in the thermodynamic limit. As a first step, we
shall find the expression for $\hat H$ on
$\hbox{S}^{2n}\setminus\{S\}$ (i.e., the sphere with the south
pole $S$ removed). To do this, we need to fix a gauge on
$\hbox{S}^{2n}\setminus\{S\}$, i.e., a smooth section $\phi$ on
$\hbox{S}^{2n}\setminus\{S\}$ of the principal bundle in
(\ref{pb}). We prefer to choose an $\hbox{SO}(2n)$-equivariant
gauge, for example, we may take
\begin{equation}\label{E:gauge}
\hbox{\bf Landau gauge}:\hskip 5pt \phi(\vec y)=\left(
\begin{matrix}
I-{2\vec y\vec y^T\over 1+y^2} & {2\vec y\over 1+y^2}\cr -{2\vec
y^T\over 1+y^2} & {1-y^2\over 1+y^2}
\end{matrix}
\right),
\end{equation}
where $\vec y$ is the standard stereographical map
$\hbox{S}^{2n}\setminus\{S\}\rightarrow {\bb{R}}^{2n}$, viewed as
a coordinate map. Then the connection form on
$\hbox{S}^{2n}\setminus \{S\}$ under the Landau gauge is
\begin{eqnarray}\label{E:conn}
\phi^*(A)=2\left(\vec yd\vec y^T-d\vec y\vec y^T\right)+ O(|\vec
y|^3),
\end{eqnarray}
where $O(|\vec y|^3)$ is a term whose coefficient in each $dy^i$
is of the order $|\vec y|^3$ as $\vec y\rightarrow \vec 0$.

For the purpose of taking limit, it is convenient to have another
representation of ${\Fr so}(2n)$ \cite{DZ73}: a complex $2n\times
2n$-matrix $X$ is in ${\Fr so}(2n)$ if and only if it is
anti-hermitian and $JX^TJ+X=0$, where $J=\left(
\begin{matrix}
0 & I_n\cr I_n & 0
\end{matrix}
\right)$ with $I_n$ being the $n\times n$-identity matrix. In this
representation, the natural connection form in equation
(\ref{E:conn}) becomes
\begin{eqnarray}
B=\left(
\begin{matrix}
zdz^\dag-dzz^\dag & zdz^T-dzz^T\cr
 & \cr
{\overline {zdz^T-dzz^T}} & {\overline {zdz^\dag-dzz^\dag}}
\end{matrix}
\right)+O(|z|^3),
\end{eqnarray}
where $z=\left(
\begin{matrix}
z^1\cr \vdots\cr z^n
\end{matrix}\right)$ with $z^\mu=y^\mu+\sqrt{-1}y^{n+\mu}$ for $1\le \mu\le n$ and
bar means complex conjugation. Then the Hamiltonian operator (in
the Landau gauge) is
\begin{equation}\label{E:H}
\hat H=-{1\over 2MR^2}{1\over \sqrt h}(\partial_j+B_j)\left[
h^{jk}\sqrt h(\partial_k+B_k)\right],
\end{equation}
where $h$ is the following Riemannian metric on
$\bb{C}^n=\bb{R}^{2n}$:
\begin{equation}
h_{ij}(\vec y)={c\over 1+{\vec y}^2}\delta_{ij},
\end{equation}
where $c$ is a constant which can/will be set to be $1$.

To find the thermodynamic limit of the Hamiltonian in equation
(\ref{E:H}), we write $B=B_\mu dz^\mu+B_{\bar\mu}d{\bar z^\mu}$,
replace $z$ by $z/\sqrt {2I}$, $\vec y$ by $\vec y/\sqrt {2I}$ and
$R^2$ by $2I$, and observe that as $I\rightarrow \infty$,
\begin{eqnarray}
B_\mu/\sqrt{2I}\rightarrow -{1\over 2}{\bar z^\mu}, \hskip 30pt
B_{\bar \mu}/\sqrt{2I}\rightarrow {1\over 2}{ z^\mu}
\end{eqnarray}
as operators, and
\begin{eqnarray}
h_{ij}\rightarrow \delta_{ij},
\end{eqnarray}
therefore,
\begin{eqnarray}
\hat H\rightarrow \hat H_\infty=\sum_{1\le \mu\le n}-{1\over
M}\left(\nabla_\mu\nabla_{\bar \mu}+\nabla_{\bar
\mu}\nabla_\mu\right),
\end{eqnarray}
where $\nabla_\mu=\partial_\mu-{1\over 2}{\overline {z^\mu}}$ and
$\nabla_{\bar \mu}=\partial_{\bar \mu}+{1\over 2} z^\mu$. In other
words, the Hamiltonian for a single charged particle in QHE on
$2n$-space is
\begin{eqnarray}\label{E:sf}
\hat H_\infty=\sum_{1\le \mu\le  n}\left (-\nabla_\mu\nabla_{\bar
\mu}+{1\over 2}\right)\hbar\omega_c.
\end{eqnarray}
Roughly speaking, equation (\ref{E:sf}) says that $\hat H_\infty$
is just the sum of $n$ copies of the Hamiltonian for a single
particle in QHE on two-space.

{\bf Main results\footnote{Before taking limit, the physics is
about a charged particle on a sphere under the influence of a
natural background gauge field. However, the limit does not have
this kind of interpretation anymore. In the semiclassical picture,
a $2n$-dimensional QHE droplet is a finite ball in the
configuration space (not ${\bb R}^{2n}$) whose Landau levels up to
the boundary are all filled. That is clear from our explicit
description of the ground-state wave-functions.} for QHE on even
dimensional spaces}. {\it Let $\hat h$ be the Hamiltonian of a
single particle in QHE on two-space in the Landau gauge, $V$ be
the Bargmann-Fock space \cite{VB62} of holomorphic functions on
$\bb{C}$. Then the Hamiltonian of a single particle in QHE on
four-space (in the Landau gauge) is
\begin{eqnarray}\label{E:ms}
\hat H_\infty= {\hat h}\otimes I\otimes I+ I\otimes {\hat
h}\otimes I,
\end{eqnarray}
and it acts on the Hilbert space  $L^2({\bb{R}}^2)\otimes
L^2({\bb{R}}^2)\otimes V$. ( here $I$ is the identity operator)
The spectrum of this Hamiltonian is
\begin{eqnarray}
E(q)=(q+1)\hbar\omega_c
\end{eqnarray}
where $q=0,1,2,\ldots$ and the Hilbert space of ground states is
the Bargmann-Fock space of holomorphic functions on $\bb{C}^3$
with orthonormal basis (when we take the magnetic length be $1$)
\begin{eqnarray}
\psi_{\bf k}(z)={z^{[{\bf k}]}\over \sqrt {\pi^{3}{\bf
k}!}}\exp{\left(-{1\over 2}|z|^2\right)},
\end{eqnarray}
where ${\bf k}\equiv (k_1,k_2,k_3)\in\bb{Z}_+^3$, $|z|^2\equiv
|z_1|^2+|z_2|^2+|z_3|^2$,  $z^{[{\bf k}]}\equiv
z_1^{k_1}z_2^{k_2}z_3^{k_3} $and ${\bf k}!\equiv k_1!k_2!k_3!$.

Similar conclusion also holds for QHE on $2n$-space with $n>2$. In
particular, the spectrum of the Hamiltonian is
\begin{eqnarray}\label{spec}
E(q)=(q+{n\over 2})\hbar\omega_c
\end{eqnarray}
where $q=0,1,2,\ldots$ and the Hilbert space of ground states is
the Bargmann-Fock space of holomorphic functions on
${\bb{C}}^{{n(n+1)\over 2}}$\,--- the configuration space}.

One may wonder whether the spectrum and the ground state
wave-functions for a single particle in QHE on $2n$-space would be
the same if they are obtained as the thermodynamic limit of their
counterparts in QHE on $2n$-spheres. The answer is yes. The
spectrum is seen to be the same by comparing equation (\ref{spec})
with equation (\ref{E:es}), and the ground state wave-functions
can be seen to be the same by doing a little further work. ( see
Appendix B)

\section{Conclusion}
The recent generalization of the QHE to four-space has been viewed
as a significant attempt to fundamental questions in physics
\cite{ZH01}, and it has attracted a lot of attentions from
physicists.  In this paper, a clean geometric construction for
this generalization is presented; and it actually yields a
sequence of models of the QHE type: \vskip 5pt
\centerline{\emph{QHE on $2$-space, Zhang-Hu Model, QHE on
$6$-space, QHE on $8$-space\footnote{In dimension 8k, if we use
real chiral spinor in our construction, the configuration space
will be smaller, its dimension will be reduced from $8k + 4k(4k-1)
= 4k(4k + 1)$ to $8k + {4k(4k-1)\over 2}=2k(4k + 3)$, i.e., from
20 to 14 when k = 1. Please compare with \cite{BHTZ03}.}, ....}}
\vskip 5pt \noindent Contrary to the accounts in \cite{BHTZ03,
ZH01}, the existence of Quantum Hall Effects does not {\it
crucially} depend on the existence of division algebras.

Moreover, the Hamiltonian and the ground state wave-functions for
a single particle on flat space are derived and explicitly
described. The simplicity of this description immediately tells us
that QHE on higher dimensional flat space shares common features
such as incompressibility with QHE on two-space. We hope the
simplicity of this description can also remove some mystery
surrounding QHE on higher dimensional spaces and thus facilitate
the search for the grand unification based on QHE on
four-space\cite{ZH01}.

\appendix
\section{Quantum mechanics of a charged particle on homogeneous spaces}

The quantum mechanics of a charged particle on homogeneous spaces
has been discussed in many recent papers in mathematical physics
journal \cite{LAND92, MT96} from more algebraic or
computational point of view.  Here we give a short presentation of
it in geometric language. No originality is claimed because the
construction is tautological and obvious to a modern geometer and
has already appeared explicitly in \cite{MS89} (at least) in the
special case.

\subsection{Generality}
Let $(\hbox{X},h)$ be a Riemannian manifold with Riemannian metric
$h$. Consider the quantum mechanics of a neutral particle of mass
$M$ freely moving in $\hbox{X}$, it is well-known that the
hamiltonian operator is
\begin{eqnarray}
\hat H={\hbar^2\over 2M}\Delta
\end{eqnarray}
where $\Delta$ is the semi-positive definite Laplace operator. (In
flat Euclidean space, $\Delta=-\sum_i\partial_i^2$) The Hilbert
space of this quantum mechanics problem is the space of square
integrable complex-valued functions on $\hbox{X}$. Following
Hodge, we write
\begin{eqnarray}
\Delta=d^\dag d
\end{eqnarray}
where $d$ is the exterior differential operator on complex-valued
functions and $d^\dag$ is its formal adjoint. In a local
coordinate system, we have
\begin{eqnarray}
\Delta=-{1\over \sqrt h}\partial_i(h^{ij}\sqrt h\partial_j)
\end{eqnarray}
where $h=\det(h_{ij})$.

Next we assume the particle is charged, and there is a background
gauge field. Geometrically, a background gauge field is just a
connection $A$ on certain hermitian vector bundle $\xi$ on
$\hbox{X}$. A connection $A$ is equivalent to linear operator
\begin{eqnarray}
d_A:\hskip 3pt \Gamma(\xi)\rightarrow \Gamma(T^*\hbox{X}\otimes
\xi)
\end{eqnarray}
satisfying the Leibnitz rule: $d_A(fs)=df\otimes s+fd_As$, where
$s$ is a section of $\xi$ and $f$ is a function on $\hbox{X}$. The
Riemannian metric on $\hbox{X}$ together with the hermitian metric
on $\xi$ makes possible the definition of the formal adjoint of
$d_A$ (denoted by $d_A^\dag$). The obvious generalization of
$\Delta$ in equation (2) is
\begin{eqnarray}
\Delta_A=d^\dag_Ad_A.
\end{eqnarray}
In a local coordinate system, with a choice of a gauge,  we have
\begin{eqnarray}
\Delta_A=-{1\over \sqrt h}(\partial_j+ A_j)\left(h^{jk}\sqrt
h(\partial_k+A_k)\right)
\end{eqnarray}
where $A_idx^i$ is the Lie algebra-valued one-form representing
the connection $A$ in the fixed gauge.

The obvious generalization  of equation (1) is
\begin{eqnarray}
\hat H={\hbar^2\over 2M}\Delta_A.
\end{eqnarray}

For a general $(\hbox{X}, h)$ and a general $(\xi, A)$, the
quantum mechanics problem is difficult to solve. However, for
homogeneous space $\hbox{X}$ and the associated canonical
$(\xi,A)$, the quantum mechanics problem is exactly soluble.

\subsection{Quantum mechanics on homogeneous space}
Let $\hbox{G}$ be a reductive Lie group, $\hbox{H}$ be a compact
Lie subgroup. The Cartan-Killing metric on $\hbox{G}$ give rise to
a canonical Riemannian metric on the homogeneous space
$\hbox{G/H}$. The principal $\hbox{H}$-bundle
\begin{eqnarray}
\hbox{H}\rightarrow \hbox{G}\rightarrow \hbox{G/H}
\end{eqnarray}
has a canonical connection:
\begin{eqnarray}
A(g)=Pr_{\hbox{\Fr h}}(g^{-1}dg)
\end{eqnarray}
where $Pr_{\hbox{\Fr h}}$ is the orthogonal projection onto {\Fr
h} (the Lie algebra of $\hbox{H}$) using the Cartan-Killing metric
on \hbox{\Fr g} (the Lie algebra of $\hbox{G}$).

Let $\hbox{V}$ be an irreducible unitary representation of
$\hbox{H}$, form the vector bundle
\begin{eqnarray}
\xi_V:\hskip 3pt \hbox{G}\times_{\hbox{H}} \hbox{V}\rightarrow
\hbox{G/H}
\end{eqnarray}

The quantum mechanics problem discussed in previous subsection,
when applied in this setting, is completely soluble. It turns out,
the problem can be fully described in terms of the representation
theory of Lie groups: The Hilbert space, being the square
integrable sections of $\xi_{\hbox{V}}$, is called the induced
representation of $\hbox{G}$ (induced from $\hbox{V}$), and the
hamiltonian is
\begin{eqnarray}
\hat H={\hbar^2\over 2M}(c_2(\hbox{G})-c_2(\hbox{H},\hbox{V}))
\end{eqnarray}
where $c_2(\hbox{G})$ is the quadratic Casimir operator of
$\hbox{G}$ and $c_2(\hbox{H},\hbox{V})$ is the value of quadratic
Casimir operator of $\hbox{H}$ on $V$. Here we have used equation
\begin{eqnarray}\label{E:m}
d_A^\dag d_A=c_2(\hbox{G})-c_2(\hbox{H},\hbox{V})
\end{eqnarray}
The proof of this equation is a simple exercise. The key
observation is that both sides commutes with the induced left
action by $\hbox{G}$ on space of $\hbox{H}$-equivariant map from
$\hbox{G}$ to $\hbox{V}$. (Note that a section of $\xi_{\hbox{V}}$
is just an $\hbox{H}$-equivariant map from $\hbox{G}$ to
$\hbox{V}$.)  Based on this observation, we just need to check
that $d_A^*d_A\phi=(c_2(\hbox{G})-c_2(\hbox{H},\hbox{V}))\phi$ at
the identity $e$ of $\hbox{G}$. Next we choose an orthonormal
basis $\{X_i\}$ at $T_e\hbox{G}$, such that the first $p$ of the
basis vectors form an orthonormal basis for the orthogonal
complement $P$ of $T_e\hbox{H}$ in $T_e\hbox{G}$, so $P\cong
{\bb{R}}^p$. Locally around $e\hbox{H}$, $\hbox{G/H}$ is
diffeomorphic to $P$ under the exponential map, so it is also
diffeomorphic to ${\bb{R}}^p$, and this defines a local coordinate
map. The next observation is that these local coordinates are
geodesic normal coordinates at $e\hbox{H}$. Then, using the
definition of covariant derivatives, we have $d^*_Ad_A
\phi|_e=\sum_{i=1}^pX_iX_i\phi|_e=c_2(\hbox{G})\phi|_e-c_2(\hbox{H})\phi|_e$.
The proof is completed by observing that
$c_2(\hbox{H})\phi|_e=c_2(\hbox{H})\cdot\phi(e)$ (here $\cdot $ is
the action of $\hbox{U}\hbox{\Fr h}$ on $\hbox{V}$).

Remark that the hamiltonian discussed in \cite{LAND92,MT96} is
\begin{eqnarray}
\hat H'=c_2(\hbox{G}),
\end{eqnarray}
and the one we use here appears explicitly in \cite{MS89} in the
case $\hbox{G}=\hbox{SU}(2)$ and $\hbox{H}=\hbox{U}(1)$.

\section{Concrete Description of wave-functions of a
single particle in QHE on even-spheres} The main purpose here is
to describe the wave-functions of a single particle in QHE on
even-spheres at a fixed energy level as certain polynomials on the
decompactified configuration space of QHE on even-spheres.

For this purpose, we choose the Landau gauge on
$\hbox{S}^{2n}\setminus\{S\}$ and identify
$\hbox{S}^{2n}\setminus\{S\}$ with $\bb{C}^n$ under the standard
stereographical projection map. Also, for each $k\ge 1$ we let
$Z_k$ denote the space of complex $k\times k$-matrices that are
skew-symmetric about the second diagonal, and identify
${\bb{C}}^k\times Z_k$ with $Z_{k+1}$ via
\begin{eqnarray}
(z,z_k) \mapsto Z\equiv\left(\begin{matrix}z &  z_k\cr 0 & -
\tilde z\end{matrix}\right).
\end{eqnarray}
Here, for $z=(z_1,\ldots,z_k)^T$---a column vector, we use $\tilde z$ to
denote $(z_k,\ldots,z_1)$---a row vector.

Note that $\Delta_I^+$ has a concrete realization as the space of
certain polynomials on $Z_n$ \cite{DZ73}. Then a smooth section of
$\xi_I$,  being uniquely specified by a smooth map from $\bb{C}^n$
to $\Delta_I^+$, can be realized as a smooth map from
$\bb{C}^n\times Z_n$ to $\bb{C}$ or a smooth map from $Z_{n+1}$ to
$\bb{C}$.

Therefore, ${\cal H}_I(q)$ can be viewed as the space of certain
smooth maps from $Z_{n+1}$ to $\bb{C}$ in a natural way. On the
other hand, ${\cal H}_I(q)\cong \Delta_I(q)$---the irreducible
representation space of $\hbox{Spin}(2n+1)$ labelled by the
highest weight state $\left |(q+I)\underbrace{I\cdots
I}_{n-1}\right >$. Since $\Delta_I(q)$ has a concrete realization
as the space of certain polynomials on $Z_{n+1}$ \cite{DZ73}, the
description of ${\cal H}_I(q)$ would be complete if we know the
natural identification ${\cal H}_I(q)\cong \Delta_I(q)$. For this
purpose, we define a self diffeomorphism $\phi$ of $Z_{n+1}$: For
$Z=\left(\begin{matrix}z & z_n\cr 0 & -\tilde z
\end{matrix}\right)$, we have
\begin{eqnarray}
\phi(Z)=\left(\begin{matrix}z & (I_n+z z^\dag)z_n(I_n+
{\tilde z}^\dag \tilde z)\cr 0 & -\tilde
z\end{matrix}\right)
\end{eqnarray}
where $I_n$ is the $n\times n$-identity matrix, and $\dag$ means
hermitian conjugation. We are now ready to state our {\bf
conclusion}: {\it The one-to-one correspondence ${\cal
H}_I(q)\ni\psi_f\leftrightarrow f\in \Delta_I(q)$ is
\begin{equation}
\psi_f(Z)=f(\phi(Z)).
\end{equation}
Moreover, the integration measure used in defining the inner
product on the space of wave-functions is
\begin{eqnarray}
d\mu_I={N_I(n)dZ\over {\left(1+|z|^2\right
)^{2I+2n}|\det(I_n+z_nz_n^\dag)|^{I+(n-1)}}}
\end{eqnarray}
where $N_I(n)$ is the normalization constant chosen such that
$\int_{Z_{n+1}} d\mu_I=1$.}

Upon replacing $Z$ by $Z/\sqrt {2I}$ and taking the limit
$I\rightarrow \infty$, we have
\begin{eqnarray}
d\mu_I\rightarrow d\mu_\infty =\pi^{-{n(n+1)\over 2}}e^{-|Z|^2}dZ,
\end{eqnarray}
i.e., the limit of the integration measure is the integration
measure of Bargmann-Fock space \cite{VB62}. It is then easy to see
that as $I\rightarrow \infty$, $\Delta_I(0)$ approaches a subspace
of the Bargmann-Fock space of holomorphic functions on
$\bb{C}^{n(n+1)/2}$, and this subspace is the whole Bargmann-Fock
space in the case $n=1$ and $n=2$. An induction argument actually
shows that this subspace is the whole Bargmann-Fock space in the
general case, too.

\end{document}